\documentclass{article}
\usepackage{dcase2025_techrep,amsmath,graphicx,url,times,booktabs, tabularx}
\usepackage{amssymb}
\usepackage{multirow}
\usepackage{pifont}

\title{JOINT FEATURE AND OUTPUT DISTILLATION FOR LOW-COMPLEXITY ACOUSTIC SCENE CLASSIFICATION}

\name{Haowen Li$^{1}$,
      Ziyi Yang$^{1}$,
      Mou Wang$^{2}$, 
      Ee-Leng Tan$^{1}$,
      Junwei Yeow$^{1}$, 
      }
\secondlinename{	  
      Santi Peksi$^{1}$, 
      Woon-Seng Gan$^{1}$
      }
\address{$^1$ Smart Nation TRANS Lab, Nanyang Technological University, Singapore\\          
        $^2$  Institute of Acoustics, Chinese Academy of Sciences, Beijing, China\\ 
        haowen.li@ntu.edu.sg, ziyi016@e.ntu.edu.sg, wangmou21@mail.nwpu.edu.cn, etanel@ntu.edu.sg,\\
        junwei004@e.ntu.edu.sg,
        speksi@ntu.edu.sg, ewsgan@ntu.edu.sg \\
 }

\begin{document}

\ninept
\maketitle

\begin{sloppy}

\begin{abstract}
This report presents a dual-level knowledge distillation framework with multi-teacher guidance for low-complexity acoustic scene classification (ASC) in DCASE2025 Task 1. We propose a distillation strategy that jointly transfers both soft logits and intermediate feature representations. Specifically, we pre-trained PaSST and CP-ResNet models as teacher models. Logits from teachers are averaged to generate soft targets, while one CP-ResNet is selected for feature-level distillation. This enables the compact student model (CP-Mobile) to capture both semantic distribution and structural information from teacher guidance. Experiments on the TAU Urban Acoustic Scenes 2022 Mobile dataset (development set) demonstrate that our submitted systems achieve up to 59.30\% accuracy.\footnote{ \url{https://github.com/HaoWLee/dcase2025_task1_inference}}

\end{abstract}

\begin{keywords}
Acoustic Scene Classification, Knowledge Distillation, Data Augmentation, Feature Distillation
\end{keywords}

\section{Introduction}
\label{sec:intro}

Acoustic Scene Classification (ASC) aims to identify the environment in which an audio recording was captured, such as a street, shopping mall, or park, based on its acoustic characteristics \cite{barchiesi2015acoustic, mesaros2018multi}. The DCASE 2025 Challenge Task 1 focuses on developing low-complexity ASC models that are robust to domain shifts across mobile recording devices and diverse urban environments. The challenge emphasizes generalization across devices under strict constraints on model size and computational cost. The task uses the TAU Urban Acoustic Scenes 2022 Mobile dataset \cite{mesaros2021taudataset}, which contains approximately 64 hours of audio recordings under 10 acoustic scenes. Compared with previous editions, the 2025 challenge emphasizes on robustness to unseen-device conditions and data efficiency. Specifically, models must be trained using only a 25\% subset of the official training set. In addition, submitted systems must not exceed 128kB parameter memory and 30M multiply-accumulate operations (MACs) per inference pass.

To address these challenges, knowledge distillation (KD) has been widely adopted to train compact student networks under the supervision of larger teacher networks~\cite{bai2024hierarchical, yeo2024teacher}. KD was first introduced by Hinton et al.\cite{hinton2015distilling} as a technique to compress large models into smaller ones by transferring soft target distributions. Since then, KD has evolved into a general learning paradigm with various forms of knowledge, including output logits, intermediate features, attention maps to relational structures \cite{gou2021knowledge}. In DCASE challenges, prior work has largely focused on output-level distillation using logits, CPJKU's submission in DCASE2023 achieved strong performance under low-complexity constraints \cite{schmid2023efficient}. However, few studies have explored feature-level supervision, which has been shown to offer additional benefits in general deep learning settings \cite{romero2015fitnets, li2021simkd}.

Models such as the Patchout Spectrogram Transformer (PaSST) \cite{koutini2021efficient} and Convolutional Patch-ResNet (CP-ResNet) \cite{koutini2021receptive} have demonstrated strong performance in ASC and served as effective teachers for compact CNN-based architectures \cite{schmid2023efficient}, and early investigations into model frameworks for ASC ~\cite{Wang2019a_t1,bai2018dcase} have also offered guidance for this work.

In this work, we employ a dual-level knowledge distillation framework that combines output-level and feature-level supervision to improve the training of a compact CP-Mobile \cite{murauer2023cpjku} student model. Specifically, we ensemble multiple high-performing teacher models (CP-ResNet and PaSST) to provide complementary guidance through soft target distributions. Inspired by advances in feature-based distillation such as FitNets \cite{romero2015fitnets} and SimKD \cite{li2021simkd}, we further align intermediate representations between a designated teacher and the student network to enhance structural transfer.

This report is organized as follows. Section~\ref{sec:data preprocessing and augmentation} describes the input feature extraction process and data augmentation techniques. Section~\ref{sec:Training and KD} details the knowledge distillation framework, including student-teacher architecture, feature matching strategies, and teacher model training procedures. Section~\ref{sec:submissions} reports the configuration and evaluation results of submitted systems for DCASE 2025 Task 1. Finally, Section~\ref{sec:conclusions} summarizes the key findings and concludes the report.

\section{Data preprocessing and augmentation}
\label{sec:data preprocessing and augmentation}

\subsection{Preprocessing}
\label{subsec:preprocessing}

All models operate on 32\,kHz audio. Log-Mel spectrograms are extracted using configurations customized for each model to balance time-frequency resolution and computational efficiency.

\begin{figure*}[t]
    \centering
    \includegraphics[width=0.95\linewidth]{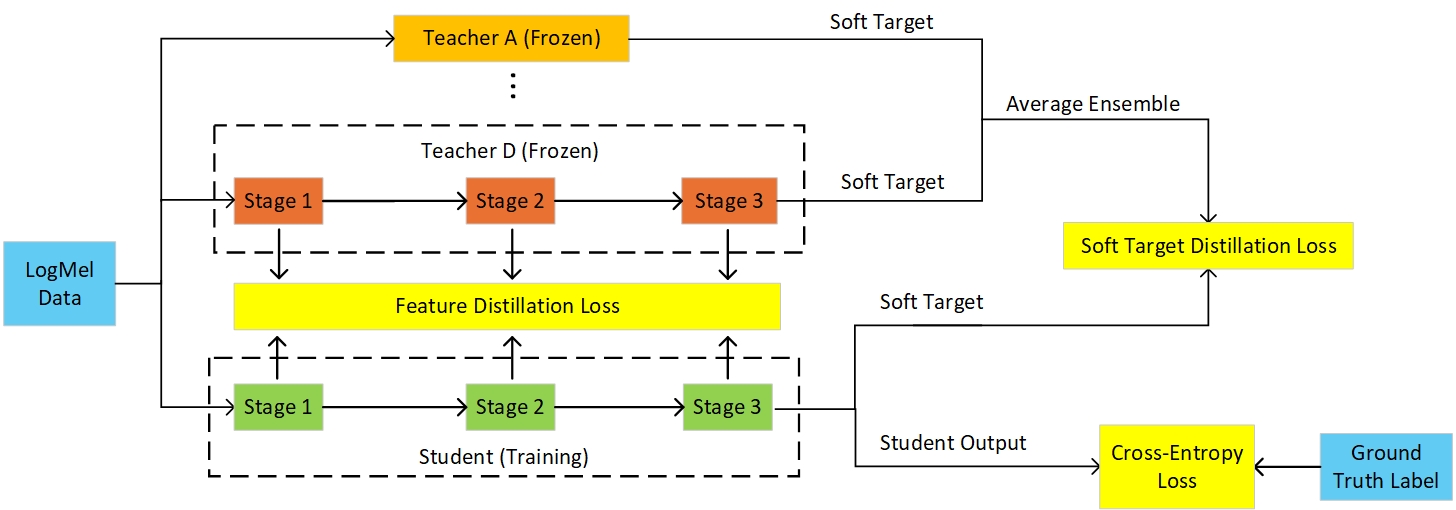}
    \caption{
    Overview of the joint feature and logit distillation framework. 
    Teacher A-D provide ensembled soft targets for output-level distillation, 
    and teacher D supervises feature-level distillation.
    }
    \label{fig:kd_architecture}
\end{figure*}

\textbf{Teacher models:} We use 2 architectures as teacher models: PaSST and CP-ResNet. For each architecture, we employ different spectrogram preprocessing settings to emphasize either frequency or temporal resolution:

\begin{itemize}
    \item \textbf{PaSST-1:} 1024-point FFT, 800-point window, 320-point hop, 128 Mel bins

    \item \textbf{PaSST-2:} 4096-point FFT, 800-point window, 320-point hop, 128 Mel bins 

    \item \textbf{CP-ResNet-1:} 4096-point FFT, 3072-point window, 750-point hop, 256 Mel bins

    \item \textbf{CP-ResNet-2:} 4096-point FFT, 3072-point window, 500-point hop, 256 Mel bins
\end{itemize}

\textbf{Student model:}
\textbf{CP-Mobile} \cite{murauer2023cpjku} uses the same 32\,kHz audio input and adopts a 4096-point FFT, 3072-point window, and a 500-point hop size, producing 256 Mel bins. Compared to CP-ResNet, this configuration provides improved temporal resolution, which aligns better with the receptive fields of lightweight models under complexity constraints.

\subsection{Data Augmentation}
\label{subsec:augmentation}

To improve model generalization under limited supervision and domain shift, three augmentation strategies are employed: time-domain rolling, frequency-domain MixStyle (Freq-MixStyle), and device impulse response (DIR) convolution.

\textbf{Time Roll:} Input waveforms are circularly shifted along the time axis to introduce temporal variability while preserving semantic content. For example, CP-Mobile and PaSST-1 apply a shift of up to 312\,ms (10,000 samples at 32\,kHz), while other models adopt a shorter shift of 125\,ms (4,000 samples).

\textbf{Freq-MixStyle:} Following the domain generalization framework MixStyle~\cite{zhou2021mixstyle}, we adopt a frequency-wise variant tailored to log-Mel spectrograms. With probability $p$, sample-wise channel statistics are interpolated using a Beta distribution with parameter $\alpha_{\text{mix}}$ to perturb style-related information and improve cross-device robustness. This augmentation is only applied to CP-ResNet teachers.

\textbf{DIR Augmentation:} To simulate the acoustic coloration introduced by different devices, waveforms are convolved with DIRs sourced from the MicIRP dataset~\cite{ micirp}. Each training example undergoes DIR-based augmentation with a specified probability to encourage invariance to microphone and channel characteristics.

The augmentation configurations for each model are summarized in Table~\ref{tab:augmentation_configs}.

\begin{table}[t]
\centering
\caption{Data augmentation configurations for teacher and student models.}
\label{tab:augmentation_configs}
\begin{tabular}{l|c|c|c}
\toprule
\textbf{Model} & \textbf{Time Roll} & \textbf{DIR Prob.} & \textbf{Freq-MixStyle} \\
\midrule
PaSST-1       & 312\,ms & 0.6 & ($\alpha_{\text{mix}}$ = 0.4,\ $p$ = 0.4) \\
PaSST-2       & 125\,ms & 0.4 & ($\alpha_{\text{mix}}$ = 0.4,\ $p$ = 0.8) \\
CP-ResNet-1   & 125\,ms & 0.4 & ($\alpha_{\text{mix}}$ = 0.4,\ $p$ = 0.8) \\
CP-ResNet-2   & 125\,ms & 0.6 & ($\alpha_{\text{mix}}$ = 0.3,\ $p$ = 0.4) \\
CP-Mobile     & 312\,ms & 0.6 & None \\
\bottomrule
\end{tabular}
\end{table}

\section{Training and knowledge distillation}
\label{sec:Training and KD}

\subsection{Teacher Model Training}
\label{subsec:teachertraining}

All teacher models are trained independently on the 25\% subset of the TAU Urban Acoustic Scenes 2022 Mobile dataset, using their respective log-Mel spectrogram configurations detailed in Section~\ref{sec:data preprocessing and augmentation}. All teachers adopt Freq-MixStyle augmentation, while DIR convolution is selectively applied to improve robustness against microphone variability.

We train 4 teacher models: 2 based on PaSST and 2 on CP-ResNet. For each architecture, one variant uses a spectrogram configuration emphasizing high frequency resolution (longer FFT and window), and the other favors higher temporal resolution (shorter window and hop size). This dual-resolution setup provides complementary time-frequency perspectives for KD.

To improve teacher model generalization and stability of soft targets, we apply model soup~\cite{wortsman2022model} within each teacher model. Specifically, we select the top 5 checkpoints after training convergence and compute the average of their weights. This simple weight averaging strategy helps mitigate overfitting and produces more robust teacher ensembles for distillation.

To generate soft targets, we compute the mean of softmax outputs from all 4 teacher models on the training set. These ensembled logits are used to supervise the student model via output-level distillation.

\subsection{Knowledge Distillation Framework}
\label{subsec:framework}

We adopt a joint knowledge distillation framework that integrates both output-level and feature-level supervision, as illustrated in Fig.~\ref{fig:kd_architecture}. The objective is to transfer knowledge from multiple high-capacity teacher models to a compact student network.

\textbf{Soft target distillation} transfers knowledge from teacher outputs. Given teacher logits $z_t$ and student logits $z_s$, we compute softened probability distributions using a temperature scaling factor $T$. The soft target loss is defined via the Kullback-Leibler divergence:
\begin{equation}
\mathcal{L}_\text{soft} = T^2 \cdot \mathrm{KL}\left( \mathrm{softmax}\left(\frac{z_s}{T}\right) ,\Vert, \mathrm{softmax}\left(\frac{z_t}{T}\right) \right).
\end{equation}

\textbf{Feature-level distillation} enforces alignment between intermediate representations, using either direct activation matching or self-similarity alignment (detailed in Section~\ref{subsec:features}). The corresponding feature loss is denoted as $\mathcal{L}_\text{feat}$.

\textbf{Cross-entropy loss} is applied between the student prediction $z_s$ and the ground-truth label $y$:
\begin{equation}
\mathcal{L}_\text{ce} = \mathrm{CE}\left(\mathrm{softmax}(z_s),, y\right).
\end{equation}

The final objective is a weighted sum of the three components:
\begin{equation}
\mathcal{L}{\text{student}} = \alpha \cdot \mathcal{L}{\text{soft}} + \beta \cdot \mathcal{L}{\text{feat}} + \gamma \cdot \mathcal{L}{\text{ce}},
\end{equation}
where $\alpha$, $\beta$, and $\gamma$ are the respective weights for soft-target, feature, and cross-entropy losses. Unless otherwise stated, we set $T = 2$, $\alpha = 1.0$, $\beta = 0.1$, and $\gamma = 0.05$ in our experiments.

\subsection{Feature Projection}
\label{subsec:features}

To bridge architectural differences between teacher and student models, we explore two distinct feature matching strategies for intermediate layer distillation: \\ 
Direct Feature Matching (DFM) aligns feature maps from teacher and student directly in the activation space~\cite{romero2015fitnets}. Specifically, we select intermediate feature maps with similar spatial dimensions and use $1{\times}1$ convolutional adapters to match channel dimensions. For example, features after the second residual block of CP-ResNet are mapped to early-stage outputs of CP-Mobile. The matching is supervised using the mean squared error (MSE) loss:
\begin{equation}
    \mathcal{L}_{\text{feat}}^{\text{DFM}} = \left\| f_s - \text{Adapter}(f_t) \right\|_2^2.
\end{equation}
This approach is inspired by conventional intermediate feature matching frameworks such as FitNets~\cite{romero2015fitnets}.

Self-Similarity Feature Matching (SSFM), adopts a self-similarity based distillation method originally proposed for speech enhancement tasks in \cite{nathoo2024two}, computes the time-frequency self-similarity Gram matrices of intermediate features for each input and minimizes the discrepancy between teacher and student similarity structures:
\begin{equation}
    \mathcal{L}_{\text{feat}}^{\text{SSFM}} = \left\| G(f_s) - G(f_t) \right\|_2^2,
\end{equation}
where $G(\cdot)$ denotes the Gram matrix capturing internal correlation across time-frequency bins.

The feature layers are manually selected based on spatial alignment and semantic consistency between teacher and student networks. We do not employ any dynamic attention or automated feature matching mechanisms in this study.

\section{Submissions and Results}
\label{sec:submissions}

We submitted systems S1 (Li\_NTU\_task1\_1) and S2 (Li\_NTU\_task1\_2) to the DCASE 2025 Task 1 evaluation. Both systems adopt CP-Mobile as the student model, trained under the proposed dual-level distillation framework. The submissions differ in their feature-level distillation strategies: S1 uses DFM, while S2 applies SSFM.

\begin{table}[t]
\centering
\caption{Detailed configuration of submissions. DFM: Direct Feature Matching, SSFM: Self-Similarity Feature Matching.}
\label{tab:submission_config_transposed}
\begin{tabular}{c|cc}
\toprule
\textbf{Submission} & \textbf{S1} & \textbf{S2} \\
\midrule
Feature KD Method          & SSFM              & DFM               \\
Feature KD Stages          & Stage 1--3        & Stage 3           \\
Feature KD Teacher         & CP-ResNet         & CP-ResNet         \\
Output KD Teacher          & \multicolumn{2}{c}{2$\times$PaSST + 2$\times$CP-ResNet} \\
Student                    & \multicolumn{2}{c}{CP-Mobile} \\
\quad \textit{Total Params}     & \multicolumn{2}{c}{61.16 K} \\
\quad \textit{MACs} & \multicolumn{2}{c}{17.05 M} \\
\textbf{Accuracy (\%)}     & 58.80             & 59.30             \\
\bottomrule
\end{tabular}
\end{table}

Table~\ref{tab:submission_config_transposed} summarizes the configuration of the submitted systems, including the feature distillation method, KD setup, and model complexity. The CP-Mobile student model used in both systems contains only 61,160 parameters and 17.05M MACs. All inference was performed using float16 precision, resulting in reduced memory usage and faster computation. The overall system design satisfies the DCASE 2025 Task 1 constraints (128kB parameter memory and 30M MACs).

S1 and S2 achieved 58.80\% and 59.30\% accuracy, respectively, indicating the potential of dual-level distillation under low-resource constraints.

\section{Conclusions}
\label{sec:conclusions}

In this report, we propose a dual-level knowledge distillation framework for low-complexity acoustic scene classification, which integrates output-level supervision from an ensemble of teacher models with intermediate feature-level guidance. To facilitate effective feature knowledge transfer, we investigate two distinct strategies: Direct Feature Matching and Self Similarity Feature Matching, with CP-ResNet employed as the feature-level teacher. All models are trained on a constrained 25\% subset of the TAU Urban Acoustic Scenes 2022 Mobile dataset. Under this setting, our submission system achieves an accuracy of up to 59.30\% on the official development set.

% -------------------------------------------------------------------------
% Either list references using the bibliography style file IEEEtran.bst
\bibliographystyle{IEEEtran}
\bibliography{refs}
%
% or list them by yourself
% \begin{thebibliography}{9}
% 
% \bibitem{dcase2016web}
%   \url{http://www.cs.tut.fi/sgn/arg/dcase2016/}.
%
% \bibitem{IEEEPDFSpec}
%   {PDF} specification for {IEEE} {X}plore$^{\textregistered}$,
%   \url{http://www.ieee.org/portal/cms_docs/pubs/confstandards/pdfs/IEEE-PDF-SpecV401.pdf}.
%
% \bibitem{PDFOpenSourceTools}
%   Creating high resolution {PDF} files for book production with 
%   open source tools, 
%   \url{http://www.grassbook.org/neteler/highres_pdf.html}.
%
% \bibitem{eWilliams1999}
% E. Williams, \emph{Fourier Acoustics: Sound Radiation and Nearfield Acoustic
%   Holography}. London, UK: Academic Press, 1999.
% 
% \bibitem{ieeecopyright}
%   \url{http://www.ieee.org/web/publications/rights/copyrightmain.html}.
%
% \bibitem{cJones2003}
% C. Jones, A. Smith, and E. Roberts, ``A sample paper in conference
%   proceedings,'' in \emph{Proc. IEEE ICASSP}, vol. II, 2003, pp. 803--806.
% 
% \bibitem{aSmith2000}
% A. Smith, C. Jones, and E. Roberts, ``A sample paper in journals,'' 
%   \emph{IEEE Trans. Signal Process.}, vol. 62, pp. 291--294, Jan. 2000.
% 
% \end{thebibliography}

\end{sloppy}
\end{document}